\documentclass[11pt]{article}
\usepackage{paper}
\usepackage{causets}

\begin{document}

%to switch ON running title
\markboth{C. Minz}{Do causal sets have symmetries?}

% \wstoc{Do causal sets have symmetries?}{C. Minz}

\title{Do causal sets have symmetries?}

\author{Christoph Minz\footnote{Institut f\"ur theoretische Physik, Universit\"at Leipzig, D-04103 Leipzig, Germany\\\url{christoph.minz@posteo.net}}}
%\aindx{Minz, C.}

\date{Contribution to the 17th Marcel Grossmann meeting (7--12 Juli 2024)}

% \address{Institut f\"ur theoretische Physik, 
% Universit\"at Leipzig\\
% D-04103 Leipzig, Germany\\
% \email{christoph.minz@itp.uni-leipzig.de}}

\maketitle

\begin{abstract}
Causal sets are locally finite, partially ordered sets (posets), which are considered as discrete models of spacetimes. 
On the one hand, causal sets corresponding to a spacetime manifold are commonly generated with a random process called sprinkling. 
This process keeps only a discrete set of points of the manifold and their causal relations (loosing the spacetime symmetries in each sprinkle).
On the other hand, the main conjecture of causal set theory is that given an ensemble of causal sets there is a corresponding spacetime manifold and the continuum symmetries of it are like all manifold properties "reconstructable" from the partial orders of all the causal sets in the ensemble. 
But most generic finite posets have very few layers ("instances of time") in contrast to sprinkles with many layers in a sufficiently large spacetime region.

In a recent project, I investigated the automorphism groups of (finite) posets in order to identify and classify their symmetries systematically. 
The comparison of local symmetries of generic posets (including Kleitmann-Rothschild orders) with sprinkled causal sets may help us to find those posets that can serve as discrete spacetime models in causal set theory.
%In the discussion of the results, we use the representation of posets as Hasse diagrams. I developed some simple tools (available for LaTeX and as interactive editor online) that will be explained alongside.
\end{abstract}

% \keywords{Causal set theory; partial orders; local symmetries}

% \bodymatter

\section{Introduction}
\label{sec:Introduction}

Causal set theory is an approach to quantum gravity that replaces spacetime manifolds by causal sets (\emph{causets} for short) \cite{BombelliLeeMeyerSorkin:1987}. 
Both causal sets and spacetime manifolds are partially ordered sets (\emph{posets}) where the partial order is given by the causal relations between events, while causets are also locally finite making the structures discrete. 
Symmetries of the causal structure are automorphisms that preserve the partial order. 
For example, the proper orthochronous Poincar\'e transformations are \emph{global} symmetries of Minkowski spacetime, where any non-trivial symmetry maps (almost) all spacetime events. 
In general, for a poset $P$ and an automorphism $\alpha \in \operatorname{Aut}( P )$, any element $a \in P$ is called \emph{fixed} if $\alpha( a ) = a$, and especially if the number of elements that are not fixed under the action of the automorphism is finite, the symmetry is \emph{local}. 

In the following, we discuss local and global symmetries in causets, in particular those that are usually considered as discrete representations of spacetime manifolds. 
At first, I recall the standard process to obtain such discrete spacetime models. 
Then I introduce local symmetries that also lead to a classification of posets and simplify their enumeration. 
Finally, I prove the absence of local symmetries in causet models of (Minkowski) spacetimes.

\section{Losing spacetime symmetries with sprinkling}
\label{sec:SpacetimeSymmetries}

In causal set theory, we replace a spacetime manifold by a discrete structure of points and their causal relations. 
A discrete representation of a given spacetime manifold $M$ is a \emph{sprinkle} that is generated by a Poisson process called \emph{sprinkling}. 
Each sprinkle is an element of the configuration space 
\begin{equation}
\label{eq:Sprinkling.ConfigurationSpace}
        \mathcal{Q}
    \coloneq \biggl\{ S \subset M \biggm\vert \forall K \underset{\mathclap{\text{compact}}}{\subseteq} M: \lvert S \cap K \rvert < \infty \biggr\}
\end{equation}
and we have a Borel $\sigma$-algbera $\mathcal{B}( \mathcal{Q} )$ of measurable subsets together with a probability measure $\mu$. 
To get an explicit expression for $\mu$, we restrict it to a (pre-)compact subset $K \subset M$ and consider the associated probability space $\bigl( \mathcal{Q}_K, \mathcal{B}( \mathcal{Q}_K ), \mu_K \bigr)$. 
The configurations $\mathcal{Q}_K$ collects all finite subsets of $K$, or equivalently, it is the union of $\mathcal{Q}_{K, n} = \{ S \subset K \mid \lvert S \rvert = n \}$ over all integer $n$. 
We remove the fat diagonal $F_n$ (that captures any duplicate elements) by defining the map $\varSigma_{K, n} : K^n \backslash F_n \to \mathcal{Q}_{K, n}$ with its pre-image denoted by $\varSigma^{-1}_{K, n}$. 
The sprinkling process in a manifold $M$ with volume measure $\nu$ at sprinkling density $\rho$ has the form 
\begin{equation}
\label{eq:Sprinkling.ProbabilityMeasure}
        \mu_K( B_n )
    = \mathrm{e}^{-\rho \nu( K )} \frac{\rho^n}{n!} \nu^n\Bigl( \varSigma^{-1}_{K, n}( B_n ) \Bigr)
\end{equation}
for any measurable subset $B_n \in \mathcal{B}( \mathcal{Q}_{K, n} )$ \cite{FewsterHawkinsMinzRejzner:2021}. 

The consequence of the discrete, random process is that the sprinkles cannot feature the symmetries of the underlying spacetime. 
Some global symmetries can only appear in very special cases of discrete structures: lattices that are invariant under discrete spatial/temporal translations and spatial rotations. 
Even though, global symmetries are lost in the sprinkling process, we usually say that sprinkles respects the spacetime symmetries, because the configuration space $\mathcal{Q}$ does contain all transformed versions of any causal set with equal probability, hence global symmetries are considered to be recoverable along with all geometric properties from the collection/ensemble.

\section{Gaining local symmetries with discreteness}
\label{sec:LocalSymmetries}

Global symmetries are not typical for causal sets, but generic causets can feature non-trivial automorphisms as \emph{local} symmetries. 
Globally hyperbolic spacetime manifolds do not have local symmetries (since the elements of any pair are distinguished by their past and future cones). 

Commonly, we represent posets with Hasse diagrams showing a vertex for each element and edges pointing up the page connecting vertices of the nearest, related elements. 
I have developed the \LaTeX~package ``causets'' (supported by the online tool ``PrOSET editor'') \cite{Minz.DiagramTools} to create Hasse diagrams in scientific writing, which is used in the following examples to visualise local symmetries. 
First we start with the simplest case. 

\begin{definition}
\label{def:ElementSymmetry}
	Let $P$ be a poset. 
	Two elements $a, b \in P$ are \emph{singleton-symmetric} if the sets of linked elements coincide, preceding (in the past, $-$) as well as succeeding (in the future, $+$), 
	\begin{equation}
	\label{eq:Symmetry.Element}
			  L^\pm( a )
		= L^\pm( b )
		\,.
	\end{equation}
\end{definition}
It is obvious from the definition that the singleton-symmetry is an equivalence relation (reflexive, symmetric and transitive).
So for any poset $P$, the quotient of $P$ by the equivalence relation inherit the partial order of $P$, which is isomorphic to a \emph{retracted} poset \cite{DuffusRival:1979} $P \oslash \pcauset[name=Singleton]{1}$. 

\begin{example}
\label{eg:ParallelSeriesComposition.Singleton}
	The quotient by the singleton-symmetry distributes over parallel ($\sqcup$) and series ($\vee$) compositions of non-singleton posets as in 
    \begin{equation}
    \label{eq:ParallelSeriesComposition.Singleton}
            \rcauset[name=ParallelSeriesExample]{14,13,12,9,15,8,11,10,5,4,2,7,1,6,3,17,16,20,19,18}{2/6}
            \oslash \pcauset[name=Singleton]{1}
        = \Biggl(
                \biggl(
                    \rcauset[name=3CrownBrokenLow]{5,4,2,7,1,6,3}{2/6}
                    \sqcup \pcauset[name=2CrownLinked3Wedge]{7,6,5,2,8,1,4,3}
                \biggr)
                \vee \pcauset[name=2Chain3Chain]{2,1,5,4,3}
            \Biggr)
            \oslash \pcauset[name=Singleton]{1}
        = \rcauset[name=ParallelSeriesExampleSingletonRetract]{10,8,11,7,9,4,2,6,1,5,3,12,13}{2/5}
        \,.
    \end{equation}
\end{example}

The poset in the example has further local symmetries. 
In general, the automorphisms that leave all elements fixed except for a finite subset constitute finite cycles in the automorphism group. 

\begin{definition}
\label{def:Symmetry.Cycles}
	Let $P$ be a poset, $Q$ be a finite poset, and $r \in \mathbb{N}$, $r \geq 2$. 
	For an automorphism $\sigma \in \operatorname{Aut}( P )$, let $\varSigma( \sigma ) \subseteq P$ denote the subset of all elements that are not fixed by $\sigma$. 
	A \emph{$( Q, r )$-generator} is an automorphism $\sigma \in \operatorname{Aut}( P )$ under the following conditions: there exists a sequence of $r$ subsets $S_i \subset \varSigma( \sigma )$ with $S_i \cong Q$, and they are the smallest, maximally ordered subsets \cite{Minz:2024} of $\varSigma( \sigma )$ with
	\begin{equation}
	\label{eq:Symmetry.SubsetCycle}
			\sigma( S_i )
		= S_{i + 1 \mod r}
		\,,
		\qquad
			\bigcup_{i = 0}^{r - 1} S_i
		= \varSigma( \sigma )
	\end{equation}
	for all $0 \leq i < r$.  
	Any distinct pair of these subsets $( S_i, S_j )$ is \emph{$( Q, r )$-symmetric}. 
	\\
	Two elements $a, b \in P$ are \emph{$( Q, r, 0 )$-symmetric} if $a = b$.
	They are \emph{$( Q, r, 1 )$-symmetric} if there exist subsets $A, B \subset P$ that are $( Q, r )$-symmetric with $( Q, r )$-generator $\sigma$ such that $a \in A$ and $b = \sigma^q( a ) \in B$ for some $1 \leq q < r$.
	Recursively setting $n = 2, 3, \dots$, the elements are \emph{$( Q, r, n )$-symmetric} if they are not $( Q, r, j )$-symmetric for any $j < n$, but there exists some $c \in P$ and $j < n$ such that $a$ is $( Q, r, j )$-symmetric to $c$ and $c$ is $( Q, r, n - j )$-symmetric to $b$.
	For short, $a$ is \emph{$( Q, r )$-symmetric} to $b$ if there exists an $n \in \mathbb{N}_0$ such that they are $( Q, r, n )$-symmetric. 
\end{definition}

Because a $( Q, r )$-generator is an automorphism $\sigma \in \operatorname{Aut}( P )$ that fixes all elements in the complement of $\varSigma( \sigma )$, any two $( Q, r )$-symmetric elements $a, b \in \varSigma( \sigma )$ also share the same (linked) preceding and succeeding elements in $P \smallsetminus \varSigma( \sigma )$, 
\begin{equation}
\label{eq:Symmetry.LinkSubsets}
		L^\pm( a ) \smallsetminus \varSigma( \sigma )
	= L^\pm( b ) \smallsetminus \varSigma( \sigma )
	\,.
\end{equation}
This is a generalisation of the singleton-symmetry and the singleton-symmetry can be seen as the special case for $Q = \pcauset[name=Singleton]{1}$ and $r = 2$. 
Similarly, the retract $P \oslash_r Q$ removes all $( Q, r )$-symmetries from $P$. 
For short, I write $P \oslash Q$ if the local symmetry is a reflection, $r = 2$. 

\begin{example}
\label{eg:ParallelSeriesComposition.2Chain}
    Note that the result of Ex.~\ref{eg:ParallelSeriesComposition.Singleton} has 2-chain-symmetries, 
    \begin{equation}
    \label{eq:ParallelSeriesComposition.2Chain}
            \rcauset[name=ParallelSeriesExampleSingletonRetract]{10,8,11,7,9,4,2,6,1,5,3,12,13}{2/5}
            \oslash \pcauset[name=2Chain]{1,2}
        = \pcauset[name=2Chain_WedgeAnd2Chain]{4,3,5,1,2,6,7}
        \,.
    \end{equation}
    The result further retracts as it has a singleton-symmetry and its retract has another 2-chain-symmetry. 
    So after retracting all symmetries, we obtain a 4-chain poset. 
\end{example}

Local symmetries and the algebraic operation $\oslash$ can be used to classify and enumerate (finite) posets \cite{Minz:2024}. 
This may also have applications to the dynamics in causal set theory \cite{Zalel:2024} and the construction of all finite posets.

\section{Losing local symmetries in infinite sprinkles}
\label{sec:SprinkleSymmetries}

Most generic causets have a small number of \emph{layers} (instances of time), which follows from the asymptotic enumeration of posets by Kleitman and Rothschild \cite{KleitmanRothschild:1975}. 
Such posets, known as Kleitman--Rothschild orders, do not model spacetime manifolds but are suppressed in the action of causal set theory \cite{LoomisCarlip:2018}. 
Kleitman--Rothschild orders can contain local symmetries more likely than sprinkles on manifolds as Minkowski spacetime, at least in the sense of the following properties.  

\begin{definition}
	For any $k \in \mathbb{N}_0$, a poset $P$ is \emph{$k$-stable locally unsymmetric} if, for every subset $S \subseteq P$ that has cardinality $0 \leq \lvert S \rvert \leq k$, the poset $P \smallsetminus S$ does not have local symmetries. 
	A poset $P$ is \emph{total locally unsymmetric} if $P \smallsetminus S$ is $k$-stable locally unsymmetric for every finite $k \leq \lvert P \rvert$. 
\end{definition}
\begin{example}
\label{eg:TotalLocallyUnsymmetric.Chains}
	Any chain poset is total locally unsymmetric. 
\end{example}

\begin{theorem}
	A sprinkle in $d$-dimensional Minkowski spacetime $\mathbb{M}^{1 + d}$ is total locally unsymmetric with probability 1. 
\end{theorem}
\begin{figure}
	\centering
	\includegraphics[scale=0.8]{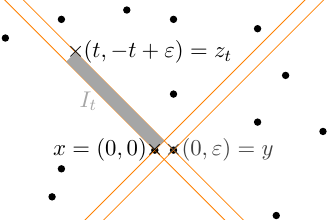}
	\caption{\label{fig:SprinkleTotalLocallyUnsymmetric} Coordinates and light cones of two points that could be (a part of) a local symmetry in a sprinkle only if the interval $I_t$ does not contain a sprinkled event (black dot) for any $t$.}
\end{figure}
\begin{proof}
    In $\mathbb{M}^{1 + 0}$, every sprinkle is a chain poset and thus total locally unsymmetric. 
    For higher dimension, we prove it by contradiction and assume that a sprinkle $\mathsf{S}$ in $\mathbb{M}^{1 + d}$ with $d > 0$ contains a pair of distinct elements $x \neq y \in \mathsf{S}$ that has the same future up to a finite set of elements $F$, and the pair becomes a local symmetry in $\mathsf{S} \smallsetminus F$. 
    We choose a coordinate system such that $x$ is at the origin and $y$ has the same coordinates except for its first spatial coordinate that is displaced by a distance $\varepsilon > 0$. 
    For any $t > 0$, let $z_t$ be the point that is displaced from $x$ by $t > 0$ in time and $- t + \varepsilon$ in the first spatial dimension. 
    An example in $\mathbb{M}^{1 + 1}$ with the event coordinates and the spacetime interval $I_t$ spanned between $x$ and $z_t$ is shown in Fig.~\ref{fig:SprinkleTotalLocallyUnsymmetric}. 
    From the sprinkling measure \eqref{eq:Sprinkling.ProbabilityMeasure}, we see that the probability for $I_t$ to contain any finite number $n$ of elements (a subset of $F$) is 
    \begin{align}
    \label{eq:Sprinkling.ProbabilityMeasure.EmptyRegion}
    		\Pr\Bigl( \lvert \mathsf{S} \cap I_t \rvert = n \Bigr)
    	&= \mathrm{e}^{-\rho \nu( I_t )} \frac{\rho^n}{n!} \nu( I_t )^n
    	\,.
    \end{align}
    The volume $\nu( I_t )$ of the compact region does not vanish since $\varepsilon$ can be arbitrarily small but is fixed to a strictly positive value, and $\nu( I_t )$ increases with $t$ linearly for $d = 1$ and more than linearly for $d > 1$ by construction. 
    Because we can choose $t$ arbitrarily large, consider $t \to \infty$, where the volume diverges and the probability~\eqref{eq:Sprinkling.ProbabilityMeasure.EmptyRegion} becomes zero for any finite $n$. 
    This implies that the region $I_t$ has to contain an infinite number of elements with probability 1, which contradicts the initial assumption that $F$ is finite. 
    So we would have to remove an infinite number of elements to make $x$ and $y$ locally symmetric. 
    Hence any random sprinkle is total locally unsymmetric with probability 1. 
\end{proof}

In summary, causet models obtained by sprinkling Minkowski spacetime do, with certainty, neither admit global nor local symmetries. 
Similar statements can be considered for many other spacetime manifolds, like those modelling the universe or black holes. 
Local symmetries and the properties of stable local unsymmetry can be exploited to distinguish between causets for discrete spacetime models and generic posets of the Kleitman--Rothschild types. 
The property of total local unsymmetry may suffice to rule out infinite posets with a finite number of layers, although this is not shown here. 
Further investigations are necessary to determine if general (total) locally unsymmetric causets are a candidate for ``manifold-like'' causets and thus relevant for applications in quantum gravity, which provides an addition to the ongoing discussion~\cite{CarlipCarlipSurya:2023,CarlipCarlipSurya:2024}.

\end{document}